\documentclass[
reprint, groupedaddress, amsmath, amssymb, aps, prb, floatfix,]{revtex4-1} 
\usepackage{color}

\usepackage{todonotes}
\usepackage{units}
\usepackage[normalem]{ulem}
\usepackage{graphicx}
\usepackage{dcolumn}
\usepackage{bm}
\usepackage{etoolbox} 
\patchcmd{\section}
  {\centering}
  {\raggedright}
  {}
  {}
\patchcmd{\subsection}
  {\centering}
  {\raggedright}
  {}
  {}

\makeatletter
\patchcmd{\NAT@citesuper}
   {\textsuperscript{#1}}{\textsuperscript{[#1]}}{}{}
\makeatother

\usepackage{array}

\usepackage{booktabs} 

\usepackage{url}
\makeatletter
\g@addto@macro{\UrlBreaks}{\UrlOrds}

\begin{document}
\title{How bright was the Big Bang?}

\author{Christopher Andersen}
\email{Electronic mail: rxz417@alumni.ku.dk} 
\author{Charlotte Amalie Rosenstroem}
\email{Electronic mail: vkc652@alumni.ku.dk}  
\author{Oleg Ruchayskiy}
\email{Electronic mail: oleg.ruchayskiy@nbi.ku.dk} 
\affiliation{Niels Bohr Institute,
University of Copenhagen, 2100 Copenhagen, Denmark}

\date{\today}

\begin{abstract}
{It is generally believed that in the epoch prior to the formation of the first stars, the Universe was completely dark (the period is therefore known as \textit{the Dark Ages}). Usually, the start of this epoch is placed at the photon decoupling. In this work, we investigate the question, whether there was enough light during the dark epoch for a human eye to see. We use the black body spectrum of the Universe to find the flux of photon energy for different temperatures and compare them with visual limits of brightness and darkness. We find that the Dark Ages actually began approximately 6 million years later than commonly stated.}
\end{abstract}

\maketitle


\section{Introduction}
The current model of the origin and the evolution of the Universe is known as the \emph{hot Big Bang} theory. 
It states that the Universe started with a period of exponentially fast acceleration known as \emph{inflation}.
After the inflation, the (largely empty) Universe was repopulated with all kinds of particles (\emph{reheating}) and continued to expand at a gradually slowing rate.
As the Universe enlarged both the density and temperature decreased, giving rise to various important processes such as the creation of asymmetry between matter and antimatter, the formation of baryons from the excess of quarks over antiquarks and the subsequent formation of light nuclei. 
All this happened while the Universe was not more than a few minutes old and had a temperature of more than a billion kelvins.\cite{EarlyUniverse,Cosmology}
At this point, electrons would still not bind to the nuclei, and hence the Universe was filled with free electrons, making it opaque to electromagnetic radiation as the photons would be scattered by the electrons. The Universe would continue to cool and after $378\,000$ years, when the temperature was about $\unit[4000]{K}$, the electrons started to combine with the nuclei to form atoms. This event is called \emph{recombination}. Shortly after the recombination, the decoupling of photons happened, where photons became free to travel through space unimpeded.\cite{EarlyUniverse}
Approximately 150 million years later the first stars began to form to lighten up the Universe.\cite{firststars,firststarsBromm}
In particular, the recent results of the EDGES experiment tell us that when the Universe was about 200 million years old, it was full of radiation provided by the first stars.\cite{NewRe}
This epoch between recombination and the formation of the first stars is commonly known as \emph{the Dark Ages}.~\cite{MiraldaEscude:2003yt} 

In this work, we investigate whether it is actually true that the Universe turned completely dark right after the decoupling of photons. 
To this end, we ``place'' a hypothetical human observer into the early Universe where no human being existed (actually, no structure more complicated than atoms of Hydrogen, Helium 
and a few other light elements existed) and analyze how much light her eye would actually register. 
We find that even without any additional optical devices, the observer would register enough photons of visual wavelengths long after photon decoupling to perceive the Universe as ``bright''.

An approach of presenting physical events through the prism of a ``human observer'' is of course not a new one. 
It is used not only in the context of popular science but also in solid scientific works.
For example, in general relativity one may discuss the following purely academic question: \emph{what would an observer see while falling into a black hole?} (see for example Refs.\cite{Blackhole1,Blackhole2,Blackhole3}). 
When discussing the future of our Universe and illustrating how a continuing accelerated expansion would look like, one again appeals to ``future cosmology'' in both research \cite{Loeb:2011cw,Loeb:2012uz,Nagamine:2003ih,Kashyap:2015lva} and popular science articles.\cite{Krauss:2007nt,Krauss:2008zza,Krauss:2010jz,2000ApJ...531...22K}

A lot of research and popular literature has been devoted to the subject of habitable planets, where one often ``depicts'' \emph{how life would look like with different kinds of suns}\cite{2017AmJPh..85...14O} including ``life in the early Universe''. \cite{Olson:2017rxj,2015Sci...350R..53S,2013Natur.504..201M,Loeb:2013fna,Loeb:2016vdd,Li:2015dsa}
Other papers about ``physics via human eyes'' can be found e.g. here \cite{Morris:1988cz,Muller:2004dq,Krauss:2001we} in part motivated by the recent movie \textit{Interstellar}. \cite{thorne2014science,Thorne,James:2015ima}

This paper takes a new step in the before-mentioned direction.
It tries to bridge a gap between an ``artistic impression'' of the history of the Universe and an actual physical picture behind it. 
It is therefore relevant to the scientific audience but also to a broader audience, interested in the subject of the early Universe.



\subsection{\label{sec:level2}The human eye}

\begin{figure}[!t]
\centering
\includegraphics[width=1\linewidth] {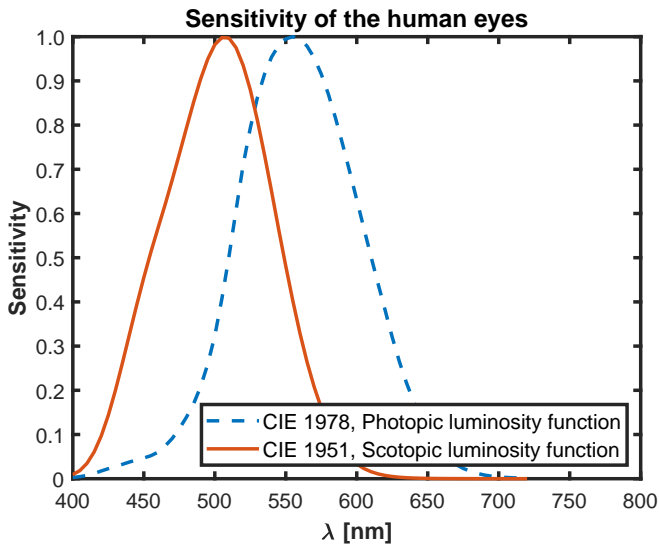}
\caption{\label{fig:sensitivity} (Color online) Sensitivity of the human eyes to light waves of different wavelengths at bright exposure (photopic vision) and dark exposure (scotopic vision). We used a \emph{photopic luminosity function} (CIE 1978 $V(\lambda) $) for the high exposures and a \emph{scotopic luminosity function} (CIE 1951 $V'(\lambda)$) for the low exposures. \cite{Humaneye}}
\end{figure}

The property of the human eye is to transform light into visual pictures in the brain. 
Light, emitted or reflected by objects, is composed of photons of different wavelengths.
The human eye is sensitive to the range of wavelengths 
\begin{equation}
\label{eq:visual}
\unit[390]{nm}\leq \lambda \leq \unit[720]{nm}, 
\end{equation}
known as the \emph{visual part of the electromagnetic spectrum}.\cite{Humaneye}
Even within the visual part of the spectrum the sensitivity of the human eye varies and the \emph{sensitivity function} (also known as a \emph{luminosity function})  describes the fraction of light registered by the eye at different frequencies. 
Additionally, different sensitivity functions apply under different light regimes. 
This is the case because two different kinds of light receptors register the photons when these hit the back of the eye.
The firsts of these are the \emph{cone cells}, which give rise to \emph{cone-mediated vision}. 
This type of vision is dominant under bright exposures and is also called \emph{photopic vision}. Secondly, we have the \emph{rod cells}, which give rise to \emph{rod-mediated vision}. 
Opposite the photopic vision, this type of vision is dominant under dark exposures and is also called \emph{scotopic vision}. \cite{Humaneye} 
A sensitivity function $V(\lambda)$ for the photopic and scotopic vision is shown in Fig.~\ref{fig:sensitivity}.
To complete our description of human vision, we note that the outer layer of the human eye, called the \emph{cornea}, works as a lens, focusing the photons towards the \emph{pupil}, which is a hole that allows the light to enter the inside of the eye. The amount of photons entering the inside of the eye is dependent on the diameter of the pupil which varies as a function of brightness. 
After entering the eye, the photons reach \emph{the double convex lens}, which again refracts the light and creates an image in the back of the eye. Finally, this image is sent towards the visual centres in the brain through nerve impulses. \cite{TheHumanEye} 


\section{\label{sec:level3} Planck's law of black body radiation and brightness estimates}

Next, we identify what one might call ``see the light''. 
We know that too much light will blind us while the opposite makes our world dark. 
Therefore, we choose two reference fluxes -- as an upper limit we use the flux from the Sun, making us blind in a short amount of time,\cite{Sunupperlimit} and as the lower limit, we use dim stars which magnitudes make them barely visible.
We assume that between those two regimes, it will thus be possible for the human eye to detect something and at the same time not be blinded by the light.

We assume throughout these calculations that all the light is in the form of the black body radiation of relic photons (that would later become the \emph{Cosmic Microwave Background}).
The flux per unit solid angle per unit wavelength is given by \emph{Planck's law} \cite{Planck} 
\begin{equation}\label{Planck's law}
B_{\lambda}(\lambda,T)=\frac{2hc^2}{\lambda^5}\frac{1}{e^{\frac{hc}{\lambda k_B T}}-1}.
\end{equation}
Here $h$ is Planck's constant, $c$ is the speed of light, $\lambda$ is the wavelength, $T$ is the black body temperature, and $k_B$ is the Boltzmann constant. 

\begin{figure}[t] 
\centering
\includegraphics[width=1\linewidth] {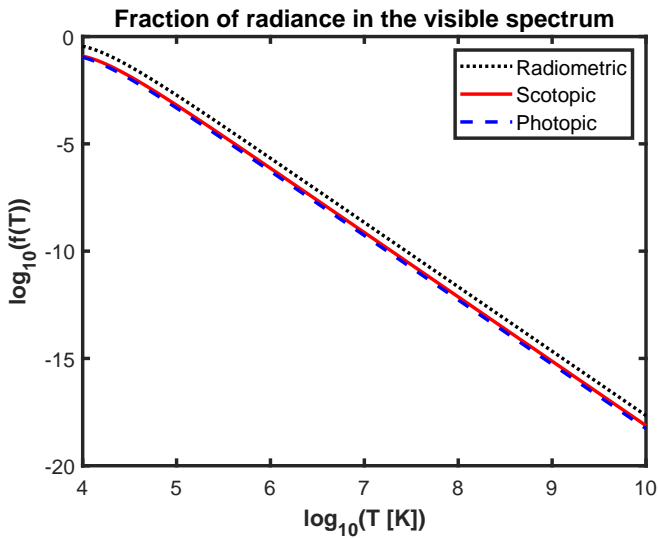}
\caption{\label{fig:fractions} (Color online) $\mathrm{log}_{10}(f(T))$ [see Eq. \eqref{eq:fraction}] for a radiometric, scotopic, and photopic vision.} 
\end{figure}

Furthermore, we assume that only visual light has any influence on the human vision, and therefore we neglect any potential damage that high-energy radiation can do to the vision/the observer. 
We thus integrate Planck's law  between the wavelengths covering the visual part of the electromagnetic spectrum, to obtain the radiance, $L_e$, 
\begin{equation}
\label{eq:radiance}
L_e(T) \equiv	 \int\limits^{\lambda_{\mathrm{max}}}_{\lambda_{\mathrm{min}}} d\lambda \, B_\lambda(\lambda,T)V(\lambda)
\end{equation}
where $\lambda_{\mathrm{max}}$ and $\lambda_{\mathrm{min}}$ are specified in Eq.~\eqref{eq:visual}, and $V(\lambda)$ is a sensitivity function. The subscript $e$ denotes 'energetic' which is a reference to \emph{radiometric quantities}. Radiometric quantities are measures of the absolute quantities of light in terms of the power of the radiation. As opposed to this, \emph{photometric quantities}, which are denoted with the subscript $v$, measure the perceived brightness of light by the human eye. \cite{Rad/Pho}
We can also define the fraction of radiance in the visual part of the spectrum as a function of the temperature. This is naturally given as
\begin{equation}\label{eq:fraction}
f(T)=\frac{\int\limits^{\lambda_{\mathrm{max}}}_{\lambda_{\mathrm{min}}} d\lambda \, B_\lambda(\lambda,T)V(\lambda)}{\int\limits^{\infty}_0 d\lambda \, B_\lambda(\lambda,T)V(\lambda)},
\end{equation}
where the denominator is just $\sigma T^4/\pi$ with $\sigma$ being the Stefan-Boltzmann constant.
Fig.~\ref{fig:fractions} shows $f(T)$ for different regimes of human vision.

\subsection{Upper limit -- the Sun}
The total flux,  received from the Sun, is  known as the Solar constant\cite{Sun2} $F_\odot = \unit[1365]{W/m^2}$.
For our purposes we should correct it by the fraction of solar energy delivered in the visible part of the spectrum, $f(T_\odot)$, where  $T_{\odot}=\unit[5780]{K}$ is the surface temperature of the Sun (atmospheric transmittance in the visible part of the spectrum is known to be close to 100\%, and therefore we neglect a corresponding correction).\cite{Sun}
Numerically, $f(T_{\odot})\simeq 0.3988$, and in this way we obtain the following flux from the Sun 
\begin{equation}
\label{Upperlimit}
F_{\mathrm{Sun,e}}=f(T_\odot) F_\odot = \unit[544.4]{W/m^2}.
\end{equation}

We will choose the value~\eqref{Upperlimit} as an \emph{upper threshold}, indicating the limit between human vision and a blindingly bright flux of light. 
\subsection{Lower limit - dim stars}

What is the dimmest object a human observer can possibly see? 
Many works have tried to estimate or measure this limit. 
The lowest possible limits are obtained when detecting glimpsing light, as described in e.g. Refs.\cite{Lowestflux,Lowestflux2,Lowestflux3} 
These estimates are not directly applicable to our case, as the Universe darkens ``slowly'', and we need to identify the dim object discernible for the human eye over longer time intervals. 
One of the ways this flux can be estimated is based on humans' ability to see dim stars. 
The limiting magnitude of a star is a statistical concept that depends upon a series of factors such as the capability to use averted vision, the individual eye sensitivity, the length of time the field has been observed etc. See the discussion in Refs.\cite{Limitingmagnitude,Limitingmagnitude2,Limitingmagnitude3} In Ref.\cite{LimitMag} the limiting magnitude is found to be 6.8, while it is meanwhile discussed that this might be excessive to the true value. As we only need a rough estimate for the calculation carried out in this paper, we choose the naked eye limiting magnitude to be 6. The flux of an object corresponding to a specific magnitude is defined as
\begin{equation}
\frac{F_1}{F_{\odot}}=100^{\frac{m_{\odot}-m_1}{5}}
\end{equation}
where $F_{\odot}$ is the solar constant, $m_{\odot}=-26.76$ is the apparent magnitude of the Sun\cite{Sun2} and $F_1$ and $m_1$ are equivalent quantities for an arbitrary star. Using the limiting magnitude $m_1=6.0$, $F_1$ is easily calculated to be $F_1=\unit[1.074 \times 10^{-10}]{W / m^2}$.
Again, using the fraction of radiance in the visual part of the spectrum,
$f(T_{\odot})$, we find the corresponding flux
\begin{equation}
\label{Lowerlimit}
F_{\mathrm{star,e}}=f(T_{\odot})F_1=\unit[4.283 \times 10^{-11}]{W / m^2}.
\end{equation}

It is instructive to convert this flux into the number of photons entering a human eye per unit time. 
Using the fact that scotopic vision has its peak sensitivity at $\lambda \simeq \unit[500]{nm}$ (Fig.~\ref{fig:sensitivity}), 
 and taking the radius of the wide-open pupil $r=\unit[4]{mm}$, 
we can approximately convert the flux~\eqref{Lowerlimit} into a number of photons via
\begin{equation}
\label{eq:photons}
N_{\mathrm{photons}} = \frac{F_{\mathrm{star,e}}\times 2 \times \pi r^2}{\frac{h c}{\unit[500]{nm}}}\approx \unit[11\times 10^3]{photons/sec}.
\end{equation}
The additional factor $2$ takes into account both eyes.
This estimate indeed reproduces the known result that a human eye can detect light, if an order of 10 photons enters the eye over a time period of $\sim \unit[10^{-3}]{sec}$.\cite{Lowestflux,Hecht2,Vavilov_UFN,Vavilov_book}
We will adopt the value~\eqref{Lowerlimit} as our lower threshold, beyond which the human eye cannot detect any light.

\begin{figure}[t] 
\centering
\includegraphics[width=1\linewidth] {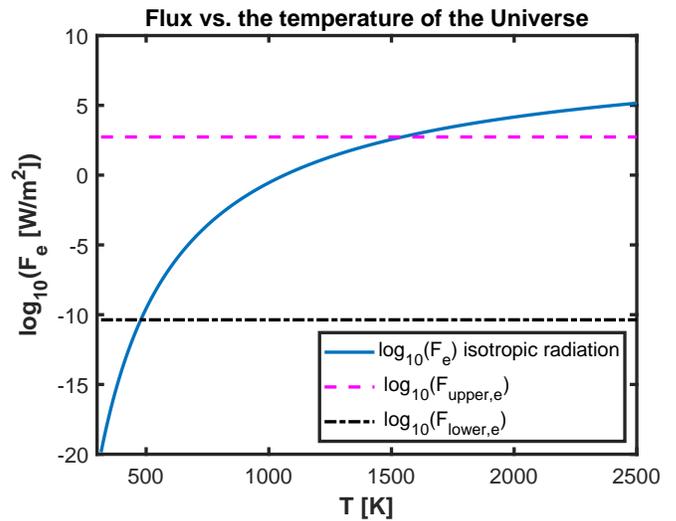}
\caption{\label{fig:rf} (Color online) flux, $F_e(T)$, inside the solid angle spanned by the human eyes, as a function of the temperature of the Universe. The upper and lower limit fluxes are plotted as reference values. The intersection between the dash-dotted and the solid line at $T=\unit[478]{K}$ marks the temperature whereby our human observer would receive the same amount of flux from the Universe and the lower limit flux. The intersection between the solid and the dashed line at $T=\unit[1544]{K}$ marks the temperature whereby this is the case for the Universe and the upper limit flux.} 
\end{figure}

\section{\label{sec:level4} Total darkness and total brightness in the Universe}

Having established the lower and upper limit for the flux, registered by the human eye, we can now determine for which temperatures the isotropic background radiation from the Universe would produce the corresponding fluxes.
For this we again use Eqs.~\eqref{Planck's law} and \eqref{eq:radiance} and scan over temperatures in order to determine a range where the flux falls between the limits~\eqref{Lowerlimit} and~\eqref{Upperlimit}. 
By mapping this range to the age of the Universe, we will be able to determine when the ``Dark Ages'' really started.

Since the radiation from the Universe is isotropic, we imagine a human sitting inside a sphere, transparent to visible light. 
Of course the human eyes do not have a field of view of $360^{\circ}$ but instead, they have an approximate field of $200^{\circ}$ horizontally and $135^{\circ}$ vertically. \cite{Fieldofview} 
We approximate this field of view by a cone with an apex angle of $160^{\circ}$.
The solid angle of such an on-axis object, having an apex angle of $2\theta$, is given by 
\begin{equation}\label{Solid angle}
\Omega=2\pi(1-\cos\theta).
\end{equation}
Using Eq.~\eqref{Solid angle} we find the solid angle covered by the human eye to be $\Omega_{\mathrm{eye}} \simeq \unit[5.2]{sr}$.
A plot of the fluxes can be seen in Fig. \ref{fig:rf}, where also the lower and upper limit fluxes are plotted as references. 

\subsection{Diffuse vs. point sources}

One may be concerned that our estimate of the limit of the total darkness is overly ``optimistic''. Indeed, it is insufficient that the eye as a whole receives the necessary amount of photons.
One needs to ensure that an individual rod receives no less than a few photons to be activated and ``register'' the light.\cite{Lowestflux,Hecht2}
Therefore, naively, instead of the field of view of the entire eye, $\Omega_{\mathrm{eye}}$, we should have used $\Omega \sim \Omega_{\rm star}$ (see e.g. the discussion in Ref.\cite{Sheffer:2018fbo}).

However, unlike a dim star, that spans a small field of view and whose image is located on a particular spot of the retina, the Universe is a \textit{homogeneous} source of light spanning the whole $\Omega_{\mathrm{eye}}$. 
As a result, different regions of the retina receive some amount of light simultaneously.
\textit{It is sufficient for any rod in any region of the retina to be activated for an observer to register the light}.
Let us estimate how many such regions exist in the eye based on the available experimental data.
For concreteness, we take an experimental setup of the experiment.\cite{Lowestflux}
In this experiment the source had an angular radius of $5'$ and about 500 rods were affected by its image on the retina (of which only a few were actually activated).\cite{Lowestflux}
We use Eq.~\eqref{Solid angle} to determine the solid angle $\Omega_{5'} \approx \unit[6.6 \times 10^{-6}]{sr}$.
The human eye has about 120 million rods.\cite{TheHumanEye}
Therefore, we can roughly estimate the ``working'' field of view as
\begin{equation}
\label{eq:rods}
\Omega_{\text{active regions}} \approx \frac{120 \times 10^6}{500} \Omega_{5'} \approx \unit[1.6]{sr}.
\end{equation}
We see that the resulting estimate is rather close to our initial $\Omega_{\rm eye}\simeq \unit[5.2]{sr}$.
Owing to the fact that we are estimating the power from the exponential tail of the Planck distribution, the difference between $\Omega_{\rm eye}$ and $\Omega_{\text{active regions}}$ gives only logarithmic correction to the temperature of the order of a few percents which justifies our treatment. 

\subsection{Radiometric limits}

By comparing the flux $F_e(T)$ to the upper limit~\eqref{Upperlimit}, we find that
the Universe with temperatures higher than $\unit[1544]{K}$ would be blindingly bright to the human eye. 
Similarly, from the comparison with the lower limit~\eqref{Lowerlimit}, we find the temperature $T\simeq \unit[478]{K}$.
For temperatures between $\unit[1544]{K}$ and $\unit[478]{K}$, it would thus be possible to see a diffuse glow, while for temperatures below $\unit[478]{K}$, the Universe would be perceived to be completely dark to the human eye. 

\subsection{Corrections for frequency-dependent sensitivity of the human eye}

Finally, we take into account that the sensitivity of the human eye changes with frequency, as discussed in Section~\ref{sec:level2}.
We convolve the sensitivity functions (Fig.~\ref{fig:sensitivity}) with the Planck curve, Eq.~\eqref{Planck's law}, to determine more correct limits of brightness and darkness for the human vision.
Doing this, we find that the temperature of a blindingly bright Universe increases to $\unit[1614]{K}$ while the temperature of a totally dark Universe would be below $\unit[518]{K}$.

\begin{table*}[!t]
  \centering
  \begin{tabular}{|r|c|c|c|c|c|c|c|}
    \hline
    & \multicolumn{3}{c|}{Brightness limit}  &\multicolumn{3}{c|}{Darkness limit} \\
    \cline{2-7}
    & Temperature & Redshift & Age of the Universe & Temperature& Redshift & Age of the Universe\\
    \hline
    Radiometric & 1544 K & 571 & $1.3\times 10^6$~yr & 478 K & 176 & $7.4\times 10^6$~yr \\
    Photometric & 1614 K & 597 & $1.2\times 10^6$~yr & 518 K & 191 & $6.5\times 10^6$~yr \\
    \hline
  \end{tabular}
  \caption{Limits of total darkness and a blindingly bright Universe. Photometric units additionally take into account the sensitivity functions of the human eyes. It is seen that the photometric modification only changes the results slightly.}
  \label{tab:results}
\end{table*}

\section{\label{sec:timeline}Timeline of the Universe}

The temperature of the Universe is related to the cosmological redshift, $z$, via \cite{Cosmology}
\begin{equation}
T=T_0(1+z). 
\end{equation}
Here $T_0$ is the present temperature of the Universe, given by $T_0=\unit[2.7]{K}$. \cite{Cosmology} 
The temperature $T=\unit[478]{K}$ corresponds to the redshift $z\simeq 176$, while the temperature of $T=\unit[1544]{K}$ corresponds to the redshift $z\simeq 571$.
At these redshifts, the expansion of the Universe is \textit{matter dominated} with negligible contributions from radiation energy density and a cosmological constant.
Therefore, one can neglect the cosmological constant $\Omega_\Lambda$ and relate the age of the Universe to its redshift via \cite{Cosmology} 
\begin{equation}
t(z)=\frac{2}{3H_0\Omega_{\mathrm{m,0}}^{1/2}(1+z)^{3/2}}. 
\end{equation}
Here $H_0$ is the \emph{current Hubble parameter}, which we take to be $H_0\simeq \unit[67.8]{km ~ s^{-1} ~ Mpc^{-1}}$. $\Omega_{\mathrm{m,0}}=0.308$ is the current \emph{density of matter}. \cite{BigBang} 
The corresponding redshifts and ages of the Universe for the temperatures of interests are shown in Table~\ref{tab:results}.

\begin{figure*}[!th]
\centering
\includegraphics[width=1\linewidth] {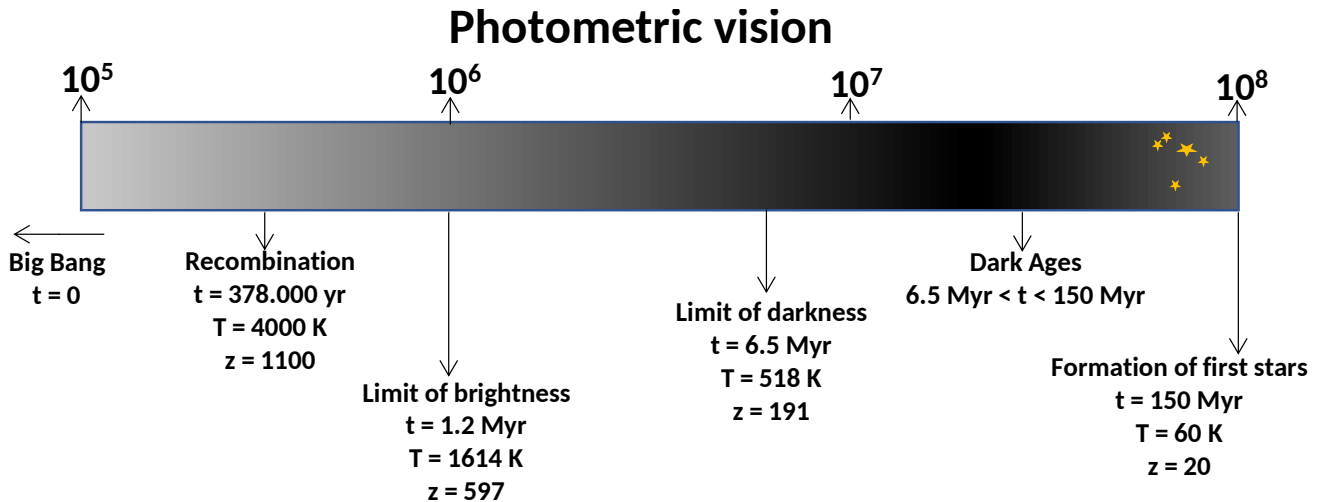}
\caption{(Color online) A logarithmic timeline of the Universe from Big Bang to the formation of the first stars. Limits of brightness and darkness (taking into account the frequency-dependent sensitivity of the human eyes), as well as the span of the Dark Ages, is shown.}
\label{fig:photometric_timeline}
\end{figure*}

We also represent our results in a timeline, describing the first 150 million years after Big Bang, in Fig. \ref{fig:photometric_timeline}.
The time of recombination, as well as the tentative epoch of the formation of first stars, is also shown, making it clear how the Dark Ages has been estimated to begin more than 6 million years later than usually stated.

\section{\label{sec:level5}Discussion}

In this paper, we analyzed the (admittedly purely academic) question about the start of the cosmological epoch known as the ``Dark Ages''. 
It is customary said that the Dark Ages began right after the hydrogen recombination was complete and photons traveled freely through the Universe.
By placing a hypothetical observer in the early Universe, and using a human eye as a proxy of the ``light detector'', we argued that the Universe did not become completely dark until it had cooled down to the temperature $T\sim \unit[500]{K}$ -- some $6$~million years after the recombination!

We also determined when the brightness of the Universe would become ``tolerable'' for the human eye. 
It turns out that this happened when the Universe cooled down to $T\sim \unit[1600]{K}$ -- about the temperature of a burning candle.

We see from the results, visualized in Fig. \ref{fig:photometric_timeline}, that the period of the Dark Ages, found in this paper, lies between photon decoupling and the formation of the first stars, which is in accordance with the general cosmological theory. 
We see that the Universe was still blindingly bright after the recombination and only became `tolerable' to the human eyes when it was $\sim 1$ million years old. 
The Dark Ages lasted from $\sim 6 -150$ million years preceded by the ``visibility window'' - when the human eye would be able to register just enough light to perceive the Universe as neither too bright nor too dim.

Throughout this paper, we have used a series of assumptions to determine the two limits of vision. These assumptions affected that only order-of-magnitude calculations were possible but at the same time made it feasible to actually carry out reasonable calculations. One of the main challenges behind human vision is that the perception of light is a complex process, in which it is close to impossible to set clear limits for the minimum amount of light needed to see something and the amount of light that would blind the human eye. This is a process which is not only effected by the amount of photon energy received per unit time but also a process influenced by internal processes in the eye, determining the amount of photon energy, actually converted into visual perception.

Finally, we note that even though more refined assumptions might have led to more precise limits,
current ambiguity about exactly how long the Dark Ages lasted, because of the uncertainty concerning when the first stars were formed, makes more precise limits more or less irrelevant. 
Hence the correction of the beginning of the ``Dark ages'' by about 6 million years (from about $400\,000$ to 6.5 million years after Big Bang) is irrelevant on the larger timescales concerning the duration of the Dark Ages but is still a major improvement in pinpointing an outset of this epoch.  

\section{Acknowledgement}

We would like to thank Steen Hansen for organizing a course that led to this project and for encouraging us to publish this article. 

\expandafter\ifx\csname url\endcsname\relax
  \def\url#1{\texttt{#1}}\fi
\expandafter\ifx\csname urlprefix\endcsname\relax\def\urlprefix{URL }\fi
\providecommand{\bibinfo}[2]{#2}
\providecommand{\eprint}[2][]{\url{#2}}

\end{document}